\documentclass{aa}  
\usepackage{graphicx}
\usepackage{txfonts}
\usepackage{natbib}
\usepackage{color}
\bibpunct{(}{)}{;}{a}{}{,}

\begin{document}
   \title{Stability and structure of analytical MHD jet formation 
          models with a finite outer disk radius}

   \author{Matthias Stute\inst{1}
	  \and
	  Kanaris Tsinganos\inst{1}
	  \and
	  Nektarios Vlahakis\inst{1}
          \and
          Titos Matsakos\inst{2}
	  \and
	  Jose Gracia\inst{3}
          }

   \offprints{Matthias Stute, \\ \email{mstute@phys.uoa.gr}}

   \institute{IASA and Section of Astrophysics, Astronomy and Mechanics, 
	Department of Physics, University of Athens, 
	Panepistimiopolis, 157 84 Zografos, Athens, Greece
	\and
	Dipartimento di Fisica Generale, 
        Universit\`a degli Studi di Torino, via Pietro Giuria 1, 
	10125 Torino, Italy
	\and
	School of Cosmic Physics, Dublin Institute of Advanced Studies, 
	31 Fitzwilliam Place, Dublin 2, Ireland
             }

   \date{Received 2 July 2008; accepted 6 September 2008}

  \abstract
   {Finite radius accretion disks are a strong candidate for launching  
     astrophysical jets from their inner parts and disk-winds are considered 
     as the basic component of such magnetically collimated outflows. 
     Numerical simulations are usually employed to answer several open questions
     regarding the origin, stability and propagation of jets. The
     inherent uncertainties, however, of the various numerical codes, applied 
     boundary conditions, grid resolution, etc., call for a parallel 
     use of analytical methods as well, whenever they are available, as a 
     tool to interpret and understand the outcome of the simulations. The only 
     available analytical MHD solutions for describing disk-driven jets are 
     those characterized by the symmetry of radial self-similarity. Those exact 
     MHD solutions are used to guide the present numerical study of disk-winds.}
   {Radially self-similar MHD models, in general, have two geometrical 
     shortcomings, a singularity at the jet axis and the non-existence of an 
     intrinsic radial scale, i.e. the jets formally extend to radial infinity. 
     Hence, numerical simulations are necessary to extend the analytical 
     solutions towards the axis and impose a physical boundary at finite radial 
     distance.}
   {We focus here on studying the effects of imposing an outer radius of the 
     underlying accreting disk (and thus also of the outflow) on the topology, 
     structure and variability of a radially self-similar analytical MHD
     solution. The initial condition consists of a hybrid of an unchanged and a
     scaled-down analytical solution, one for the jet and the other for its 
     environment.}
   {In all studied cases, we find at the end steady two-component solutions. The
    boundary between both solutions is always shifted towards the solution
    with reduced quantities. Especially, the reduced thermal and magnetic
    pressures change the perpendicular force balance at the ``surface'' of the
    flow. In the models where the scaled-down analytical solution is
    outside the unchanged one, the inside solution converges to a solution with 
    different parameters. In the models where the scaled-down analytical 
    solution is inside the unchanged one, the whole two-component solution 
    changes dramatically to support the flow from collapsing totally to the 
    symmetry axis.}
   {It is thus concluded that truncated exact MHD disk-wind solutions which may 
   describe observed jets associated with finite radius accretion disks, are 
   topologically stable.}


   \maketitle

\section{Introduction}

Astrophysical jets are ubiquitous, occurring in a variety of objects on very 
different size and mass scales. They can be produced by pre-main sequence stars 
in Young Stellar Objects, by post-AGB stars in pre-planetary (PPNs) and 
planetary nebulae (PNs), by white dwarfs (WDs) in supersoft X-ray sources and 
symbiotic stars (SySs), by neutron stars in X-ray Binaries, by stellar black 
holes in Black Hole X-ray Binaries and by supermassive black holes in the case 
of Active Galactic Nuclei. However, the formation mechanism of jets remains 
very poorly understood. 

In all such cases, jets and disks seem to be inter-related. Disks provide the
jets with the ejected plasma and magnetic fields and jets are possibly the 
most efficient means to remove excess angular momentum in the disk 
\citep[e.g.][]{Fer07} making accretion possible in the first place. 
Observationally, such a correlation is now already well established 
\citep[e.g.][in the case of YSOs]{HEG95}. Hence, our current understanding is 
that jets are fed by the material of an accretion disk surrounding the central
object. The gravitational energy of the infalling material in the disk is 
converted into kinetic energy of the outflowing matter. Radiative launching 
can be excluded due to the involved small radiation fields which have usually 
luminosities of only a few percent of the necessary kinetic luminosities 
\citep[e.g.][in the case of YSOs]{Lad85}. Furthermore, the kinetic luminosity 
of the outflow seems to be a large fraction of the rate at which energy is 
released by accretion. With such a high ejection efficiency it is natural to 
assume that jets are driven magnetically from an accretion disk; the magnetic 
model of a disk-wind seems to explain simultaneously acceleration, collimation 
as well as the observed high jet speeds \citep{KoP00}.

In their seminal analytical work \citet{BlP82} studied the magneto-centrifugal 
acceleration along magnetic field lines threading an accretion disk. They were 
the first to show the braking of matter in the azimuthal direction inside the 
disk and the outflow acceleration above the disk surface guided by the poloidal 
magnetic field components. Toroidal components of the magnetic field then 
collimate the flow. Numerous semi-analytic models extended the work of 
\citet{BlP82} along the guidelines of radially self-similar solutions of the 
full magnetohydrodynamics (MHD) equations \citep{VlT98}. 
\citet{OgL98,OgL01} solved for the local vertical structure of a thin disk 
threaded by a poloidal magnetic field of dipolar symmetry. They showed that jet 
launching occurs only for a limited range of field strengths and a limited range
of inclination angles. 

The model of \citet{BlP82}, however, has basically two serious limitations. 
First, the outflow speed at large distances does not cross the corresponding 
limiting characteristic, with the result that the terminal wind solution is 
not causally disconnected from the disk. Later, \citet[][V00 hereafter]{VTS00} 
showed that a terminal wind solution can be constructed which is causally 
disconnected from the disk and hence any perturbation downstream of the 
superfast transition cannot affect the whole structure of the steady outflow. 
The second limitation of self-similar models in general is their geometrical 
nature. Singularities exists at the jet axis in radial self-similar models. 
Numerical simulations are necessary to extend the analytical solutions, as it 
has been already done in \citet[][GVT06 hereafter]{GVT06} and \citet[][M08 
hereafter]{MTV08}. 

Another geometrical limitation of radially self-similar models is the 
non-existence of an intrinsic scale. The jets formally extend to radial 
infinity. The aim of this paper is to investigate numerically, how imposing an 
outer radius of the jet, i.e. cutting off the analytical solution at arbitrary 
radii, affects the topology, structure and stability of a particular radial 
self-similar analytical solution and hence its ability to explain observations.

Several other numerical studies exist, which have focused on the magnetic 
launching of disk-winds. In most models a polytropic equilibrium accretion disk 
was regarded as a boundary condition 
\citep[e.g.][]{UKR95, KLB99, KLB03, OCP03, NaM04, ALK05, ALK06, PRO06}. The 
magnetic feedback on the disk structure is therefore not calculated 
self-consistently. Only in recent years were the first simulations including 
the accretion disk self-consistently in the calculations of jet formation 
presented \citep[e.g.][]{CaK02, CaK04, KMS04, ZFR07}. Numerical simulations of 
stellar winds have also been done by e.g. \citet{MaP05a,MaP05b}. However, in 
none of these studies the disk has been truncated to study the effects of an 
intrinsic scale.

This paper is organized as follows: in Sec.~\ref{sec_selfsim}, the analytical 
self-similar model underlying our numerical study is reviewed. The numerical 
setup is described in Sec.~\ref{sec_num_models}. The study starts with several 
test simulations which are described in detail and whose results are presented 
in Sec.~\ref{sec_testsims}. In Sec.~\ref{sec_science_runs}, is described the 
parameter study of steady solutions and its results. We close with a summary 
of the results in the last section.

\section{Analytical self-similar model} \label{sec_selfsim}

The ideal time-dependent MHD equations which are solved numerically are:
\begin{eqnarray}
\frac{\partial \rho}{\partial t} + \nabla\cdot(\rho\,\vec v) &=& 0 \,, \\
\frac{\partial \vec v}{\partial t} + (\vec v \cdot \nabla)\,\vec v + 
\frac{1}{\rho}\,\vec B\times(\nabla\times\vec B) + \frac{1}{\rho}\,\nabla\,p 
&=& -\nabla\Phi \,, \\
\label{energy}
\frac{\partial p}{\partial t} + \vec v\cdot\nabla\,p + 
\Gamma\,p\,\nabla\cdot\vec v &=&
\Lambda \,, \\
\frac{\partial \vec B}{\partial t} - \nabla\times(\vec v\times\vec B) &=& 0
\,, \\ \nabla\cdot \vec B &=& 0 \,,
\end{eqnarray}
where $\rho$, $p$, $\vec v$, $\vec B$ denote the density, pressure, velocity 
and magnetic field over $\sqrt{4\,\pi}$, respectively. $\Phi = -\mathcal{G}\,
\mathcal{M} / r$ is the gravitational potential of the central object with mass
$\mathcal{M}$, $\Lambda$ represents the volumetric energy gain/loss terms 
($\Lambda = [\Gamma - 1]\,\rho\,Q$, with $Q$ the energy source terms per unit 
mass), and $\Gamma$ is the ratio of the specific heats. The spherical radius 
is denoted by $r$ and the cylindrical radius by $R$.

By assuming steady-state and axisymmetry, several conserved quantities exist 
along the field lines \citep{Tsi82}. By introducing the magnetic flux function 
$A = ( 1 / 2\,\pi ) \int \vec B_p \cdot d\vec S$ to label the iso-surfaces 
that enclose constant poloidal magnetic flux, then these integrals take the 
following simple form:
\begin{eqnarray} \label{integrals1}
\Psi_A (A) &=& \frac{\rho\,v_p}{B_p} \,, \\
\Omega (A) &=& \frac{1}{R}\,\left( v_\phi - \frac{\Psi_A\,B_\phi}{\rho} \right)
\,, \\
L (A) &=& R\,\left( v_\phi - \frac{B_\phi}{\Psi_A} \right) \,,
\end{eqnarray} 
where $\Psi_A$ is the mass-to-magnetic-flux ratio, $\Omega$ the field angular
velocity, and $L$ the total specific angular momentum. The ratio 
$\sqrt{L / \Omega}$ defines the Alfv\'enic lever arm $R_{\rm A}$ at the point 
where the flow speed is equal to the poloidal Alfv\'enic one. At the reference 
field line (see below) we set $R_{\rm A}|_{\alpha=1} = R_*$. In the 
adiabatic-isentropic case where $\Lambda = 0$, there exist two more integrals, 
the ratio of total energy flux to mass flux $E$ and the specific entropy $Q$, 
which are given by:
\begin{eqnarray} \label{integrals2}
E (A) &=& \frac{v^2}{2} + \frac{\Gamma}{\Gamma - 1}\,\frac{p}{\rho} + \Phi - 
\Omega\,R\,\frac{B_\phi}{\Psi_A} \,, \\
Q (A) &=& \frac{p}{\rho^\Gamma} \,.
\end{eqnarray} 

We use the steady, radially self-similar solution which is described in 
V00 and crosses all three critical surfaces. We note 
that a polytropic relation between the density and the pressure is assumed, 
i.e. $P = Q(A)\,\rho^\gamma$, with $\gamma$ being the effective polytropic 
index. Equivalently, the source term in Eq.~(\ref{energy}) has the special 
form 
\begin{equation}
\Lambda = (\Gamma - \gamma)\,p\,(\nabla\cdot\vec v) \,,
\end{equation}
transforming the energy Eq.~(\ref{energy}) to
\begin{equation}
\frac{\partial p}{\partial t} + \vec v\cdot\nabla\,p + 
\gamma\,p\,\nabla\cdot\vec v = 0 \,.
\end{equation}

The latter can be interpreted as describing the adiabatic evolution of a gas 
with ratio of specific heats $\gamma$, whose entropy $P / \rho^\gamma$ is 
conserved along each streamline.

The solution is provided by the values of the key functions $M(\theta)$, 
$G(\theta)$ and $\psi(\theta)$, which are the Alfv\'enic Mach number, the 
cylindrical distance at any point on a field line
in units of the corresponding Alfv\'enic lever arm and the 
angle between a particular poloidal field line and the cylindrical radial 
direction, respectively. $\theta$ is the colatitude in a spherical coordinate
system. Then, the poloidal field lines can be labeled by
\begin{equation} \label{eq_alpha}
A = \frac{B_*\,R_*^2}{x}\,\alpha^{x/2}, \qquad \textrm{with} \qquad \alpha = 
\frac{R^2}{R_*^2\,G^2} \,.
\end{equation}

The physical variables can be constructed as follows (V00)
\begin{eqnarray}
\rho &=& \rho_*\,\alpha^{x-3/2}\,\frac{1}{M^2} \,, \\
p &=& p_*\,\alpha^{x-2}\,\frac{1}{M^{2\,\gamma}} \,, \\
v_p &=& - v_*\,\alpha^{-1/4}\,\frac{M^2}{G^2}\,\frac{\sin\theta}{
\cos(\psi + \theta)}\,(\cos\psi\,\vec e_R + \sin\psi\,\vec e_z) \,, \\
v_\phi &=& v_*\,\lambda\,\alpha^{-1/4}\frac{G^2-M^2}{G\,(1-M^2)} \,, \\
B_p &=& -B_*\,\alpha^{x/2-1}\,\frac{1}{G^2}\,\frac{\sin\theta}{
\cos(\psi + \theta)}\,(\cos\psi\,\vec e_R + \sin\psi\,\vec e_z) \,, \\
B_\phi &=& -B_*\,\lambda\,\alpha^{x/2-1}\,\frac{1-G^2}{G\,(1-M^2)} \,.
\end{eqnarray}
Note that $x$ is a model parameter governing the scaling of the magnetic field 
and is related to $\xi = 2\,(x - 3/4)$ which is a local measure of the disk 
ejection efficiency in the model of \citet{Fer97}. The starred quantities are 
related to their characteristic values at the Alfv\'en radius $R_*$ along the 
reference field line $\alpha = 1$. Moreover, they are interconnected with the 
following relations:
\begin{equation} \label{norm_factors}
v_* = \frac{B_*}{\sqrt{\rho_*}} \,, \qquad p_* = \frac{\beta\,B_*^2}{2} \,, 
\qquad \mathcal{K} = \sqrt{\frac{\mathcal{G}\,\mathcal{M}}{R_*\,v_*^2}} \,.
\end{equation}
The constants $\lambda$ and $\mathcal{K}$ are the specific angular 
momentum of the flow in units of $v_*\,R_*$ of the reference field line and the 
Keplerian velocity at distance $R_*$ measured on the disk in units of $v_*$, 
respectively. Finally $\beta$ is the ratio of gas pressure to magnetic pressure 
and is proportional to the gas entropy.

\section{Numerical setup} \label{sec_num_models}

We solve the MHD equations with the 
PLUTO\footnote{publicly available at http://plutocode.to.astro.it/} code 
\citep{MBM07}, a modular Godunov-type code particularly oriented towards the 
treatment of astrophysical flows in the presence of discontinuities. For the 
present case, second order accuracy is achieved using a Runge-Kutta scheme (for 
temporal integration) and piecewise linear reconstruction (in space). All the 
computations were carried out with the simple (and computationally efficient) 
Lax-Friedrichs solver.

We define the reference length $R_*$ to be unity, while the reference velocity 
is normalized by setting $v_* = 1$. Time is given in units of 
$t_0 = 2\,\pi\,\sqrt{R_*^3 / \mathcal{G}\,\mathcal{M}}$, i.e. one Keplerian 
orbit at $R_* = 1$. The model parameters of this solution were chosen as 
$x = 0.75$ and $\gamma = 1.05$, while the solution parameters are given to be 
$\lambda =11.70$, $\beta = 2.99$, $\mathcal{K} = 2.00$, corresponding to the 
solution in V00 crossing all critical points. At the symmetry axis, 
the analytical solution is modified as described in GVT06 and M08.

To study the influence of the truncation of the analytical solution, we divide
our computational domain into a jet region and an external region, which are
separated by a truncation field line $\alpha_{\rm trunc}$. For smaller values 
of $\alpha$ -- i.e. smaller radii -- our initial conditions are fully 
determined by the solution of V00. The way the external region should
be initialized, however, is not as obvious. 

Therefore the technical questions, which arise at this point and which are 
tested in several simulations described below, are the following:
\begin{itemize}
\item What are the consequences of different initializations of the external
region?
\item How is the choice of $\alpha_{\rm trunc}$ affecting the results?
\item What are the effects of numerical resolution and domain size.
\end{itemize}

We impose axisymmetry at the axis, i.e. variables are symmetric across the 
boundary, normal components of vector fields and also angular components change 
sign there. Outflow conditions are set at the top $z$ and outer radial 
boundaries, i.e. all gradients across these boundaries are set to zero. At the 
lower boundary, we keep the quantities {\it fixed} to their analytical values, 
however, making sure that the problem is not over-specified. Thus, the number of
quantities set by the analytical solution is lowered by one for each critical 
surface already crossed upstream of the boundary (for details, see M08). At 
small radii, where the flow is super-fast-magnetosonic, seven of the eight MHD 
quantities are given by the analytical solution, $v_R$ is set by the requirement
$v_P \parallel B_p$. At larger radii, where the flow is sub-fast and 
super-Alfv\'enic, only six quantities are fixed, $B_\phi$ is set by symmetrizing
the values inside the domain across the boundary. In the sub-Alfv\'enic and 
super-slow regime, also $p$ is given by its values inside the domain, and in the
sub-slow regime, also $B_R$ is given.

\section{Test simulations} \label{sec_testsims}

\subsection{Initial conditions}

To answer the questions posed above, we have run ten models, together with 
the unchanged model of V00, hereafter labelled ADO ({\em analytical 
disk outflow solution}, as in M08). Details of these test models are given 
in Table \ref{tbl_test_models}. In all these models the internal part of the
flow is initialized with the V00 solution. We probe the second question by 
comparing the models FL1 -- FL3, where the analytical solution is truncated at 
$\alpha_{\rm trunc} = 0.4$, $\alpha_{\rm trunc} = 0.8$ and 
$\alpha_{\rm trunc} = 5$, respectively. In these models the external region is 
initialized with constant values, the values of the analytical solution at the 
point ($R=R_{\rm max}$, $Z=Z_{\rm max}$). The influence of numerical resolution 
and domain size is tested with the models CD1 -- CD5, in which
the external region is initialized using the analytical solution, but with
$\rho$ and $v_z$ damped with $\alpha_{\rm trunc}/\alpha$. The first and most 
important question can be addressed with models FL1, CD1 and ER1 -- ER2.
In model ER1 we use the analytical solution in the external domain, but
$v_z$ is set to a smaller value. In model ER2 all quantities are damped 
with an exponential factor in the external domain.

\begin{table*}
\caption{List of numerical test models}
\label{tbl_test_models}
\centering
\begin{tabular}{l l l l l l}
\hline\hline
Name & Grid size & Resolution & $t_{\rm end}$ & $t_{\rm coll}$ & Description \\
\hline
ADO  & [0,50] $\times$ [6,100]   & 200 $\times$ 400  & 50.0 & --   & 
model of \citet{VTS00}, \citet{GVT06} and \citet{MTV08} \\
\hline
\multicolumn{6}{c}{Models to test the influence of different 
truncation field lines} \\
\hline
FL1  & [0,50] $\times$ [6,100]   & 200 $\times$ 400  & 8.8  & 5.7  & 
$\alpha_{\rm trunc} = 0.4$, constant values in external region 
(analytical values at $[R_{\rm max},Z_{\rm max}]$) \\
FL2  & [0,50] $\times$ [6,100]   & 200 $\times$ 400  & 7.3  & 6.7  & 
$\alpha_{\rm trunc} = 0.8$, constant values in external region 
(analytical values at $[R_{\rm max},Z_{\rm max}]$) \\
FL3  & [0,50] $\times$ [6,100]   & 200 $\times$ 400  & 10.0 & 8.1  & 
$\alpha_{\rm trunc} = 5.$, constant values in external region 
(analytical values at $[R_{\rm max},Z_{\rm max}]$) \\
\hline
\multicolumn{6}{c}{Models to test the influence of the computational 
domain} \\
\hline
CD1  & [0,50] $\times$ [6,100]   & 200 $\times$ 400  & 6.3  & 2.7  & 
$\alpha_{\rm trunc} = 0.4$, in external region damping of $\rho$ and 
$v_z$ with $\alpha_{\rm trunc}/\alpha$ \\
CD2  & [0,100] $\times$ [6,200]  & 200 $\times$ 400  & 8.9  & 3.8  & 
same as CD1, but with larger domain (lower resolution, same grid size) \\
CD3  & [0,100] $\times$ [6,200]  & 400 $\times$ 800  & 8.1  & 4.1  & 
same as CD1, but with larger domain (same resolution) \\
CD4  & [0,200] $\times$ [6,400]  & 400 $\times$ 800  & 9.8  & 6.9  & 
same as CD2, but with even larger domain \\
CD5  & [0,400] $\times$ [6,100]  & 1600 $\times$ 400 & 4.7  & 4.5  & 
same as CD1, but with eight times larger domain in $R$ direction, same res. \\
\hline
\multicolumn{6}{c}{Models to test different initializations of the 
external region} \\
\hline
ER1  & [0,50] $\times$ [6,100]   & 200 $\times$ 400  & 50.0 & 24.9 & 
$\alpha_{\rm trunc} = 0.4$, in external region damping of $v_z$ to 
constant value \\ & & & & & (10\% of the value at the truncation field line) \\
ER2  & [0,50] $\times$ [6,100]   & 200 $\times$ 400  & 3.6  & --   & 
$\alpha_{\rm trunc} = 0.4$, in external region damping of all variables 
with $\exp [- (\alpha/\alpha_{\rm trunc})^2 ]$ \\
\hline
\end{tabular}
\end{table*}

\subsection{Results of the test simulations}

\begin{figure*}[ht!]
  \includegraphics[width=\textwidth]{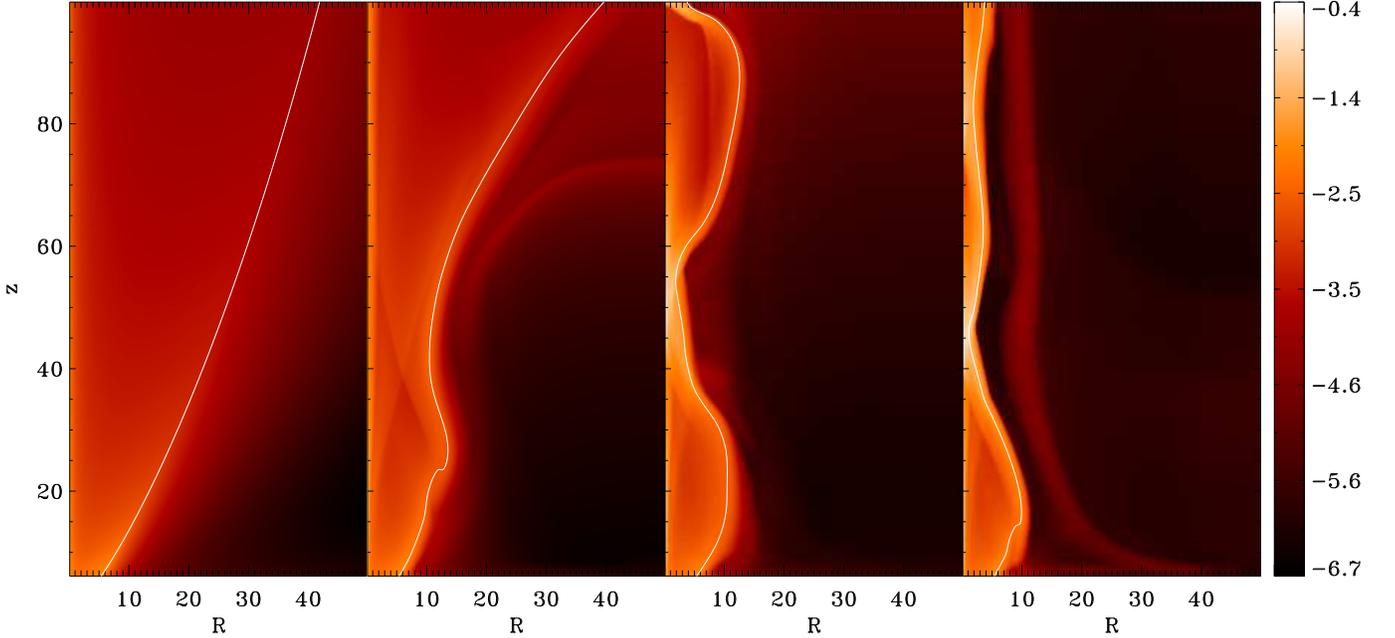}
  \caption{Structure of the flow (logarithmic density plots) for model CD1 at 
    timesteps $t = 0, 2, 3, 6\,t_0$, respectively. This is an example for the 
   collapse occurring in all test runs except for model ER2. The truncation 
   field line (the magnetic field line with $\alpha = \alpha_{\rm trunc} = 0.4$ 
   which is anchored in the lower boundary at $R = 5.375$) is also plotted 
   (white line).}
  \label{Fig_struct_modelCD1}
\end{figure*}

\begin{figure*}[ht!]
  \includegraphics[width=\textwidth]{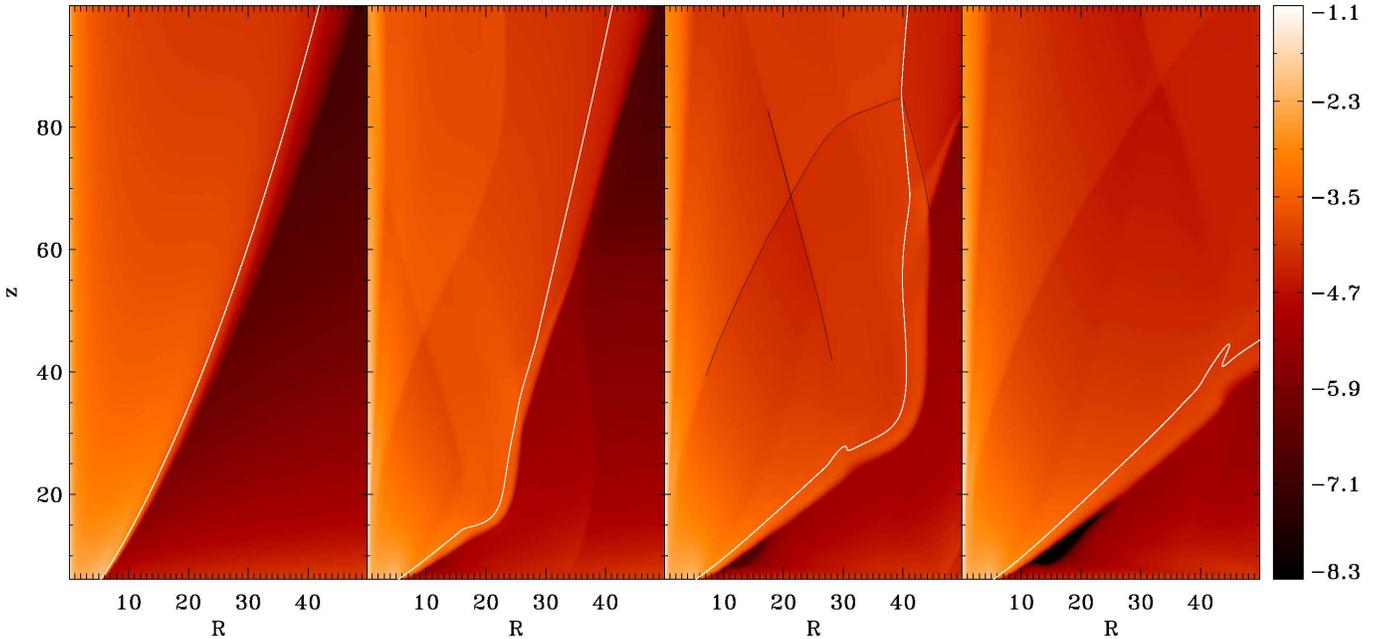}
  \caption{Structure of the flow (logarithmic density plots) for model ER2 at 
    timesteps $t = 0, 1, 2, 3\,t_0$, respectively. This is the only model 
    without a collapse. The white line is the truncation field line 
    (the magnetic field line with $\alpha = \alpha_{\rm trunc} = 0.4$ which is 
    anchored in the lower boundary at $R = 5.375$). The positions of the slow- 
    and fast-magnetosonic waves emitted by the truncation field line and of the 
    fast magnetosonic separatrix surface (FMSS) are marked by black lines.}
  \label{Fig_struct_modelER2}
\end{figure*}

The basic evolution is similar in most models -- as an example we show the 
structure of the flow for models CD1 and ER2, see 
Figs.~\ref{Fig_struct_modelCD1} and~\ref{Fig_struct_modelER2}) -- and can be 
divided into four phases: 
\begin{enumerate} 
\item At the beginning, a shock front starting at the jet base runs across the 
jet, bending its outer surface and forming a dent which then travels out- and 
upwards (see Fig.~\ref{Fig_struct_modelER2}, third panel, black lines). The 
front consists of two distinct shocks, presumably slow- and fast-magnetosonic 
waves which transport downstream the effect of the boundary condition at the 
base, namely the truncation of the solution. They can be seen in all models,
but are not present in the analytical reference model ADO as expected. A third
feature, which is also present in model ADO, is the fast 
magnetosonic separatrix surface (FMSS) which shields the flow from the
modification close to the axis (GVT06, M08).
\item Behind the dent, a new smooth jet surface develops, sometimes with a 
larger radius than the initial one. This configuration is stable for several 
$t_0$ in all models.
\item In the next phase, the outer medium prevails and compresses the jet 
along its full length, pinching it even close to the rotation axis at 
certain $z$ values. Here very dense knots are formed. We call this phase the 
collapse of the solution, since the jet width becomes smaller than the outer 
disk radius (e.g. 5.375, if $\alpha_{\rm trunc} = 0.4$). The only model without 
this collapse is model ER2.
\item Finally, jet material pushes back and a more or less cylindrical 
topology is established. The jet pulsates along the whole jet length in the 
computational domain and the time step decreases by three up to nine orders of 
magnitudes, making it impossible to further advance the simulations.
\end{enumerate}


The fact that a new smooth jet surface forms right behind the dent is 
a promising result, which may be a hint for a new equilibrium and thus a 
steady solution. However, this couldn't be tested in these runs, since the
solution collapsed. 

\begin{figure}[ht!]
  \centering
  \includegraphics[width=\columnwidth]{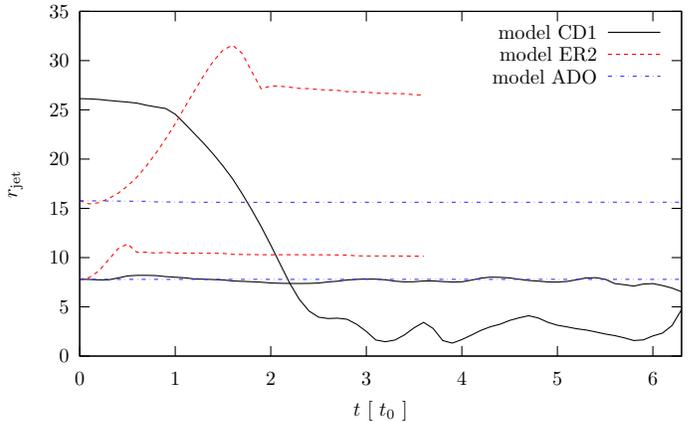}
  \caption{Evolution of the jet radius (radius of the truncation field line)
    for models CD1, ER2 and also the unchanged analytical solution ADO, at two 
    representative $z$ values above the equatorial plane ($z = 10$ and $z = 25$ 
    in the models ER2 and ADO; $z = 10$ and $z = 50$ in the model CD1). The 
    models ADO and ER2 are stationary while the model CD1 collapses.}
  \label{Fig_jetradii_testruns}
\end{figure}

In Fig.~\ref{Fig_jetradii_testruns} we plot the radii of the jets in  
models CD1, ER2 and ADO at several values of $z$. These radii correspond to 
the truncation field line (anchored in the lower boundary where 
$\alpha = \alpha_{\rm trunc}$ at $R = 5.375$), although this is not the 
outermost field line inside the jet for all timesteps in all models.

In model CD1, the jet radius stays constant until the collapse. The collapse 
starts at high latitudes (about $z = 50$) and then affects the jet further 
down-stream. After the collapse, the radius varies along the jet length and 
also with time. A similar structure is also seen in models CD2--CD4 and 
FL1--FL3 with similar values of the jet radii.

The time $t_{\rm coll}$ at which the jet collapses is different in each model
(Table \ref{tbl_test_models}). In models CD1--CD4 testing the influence of 
numerical resolution and domain size we can clearly see that the size 
of the computational domain seems to be important for triggering the collapse.
The larger the computational domain, the later the collapse occurs. The 
similarity between models CD2 and CD3 shows that the resulting timescale 
does not depend on the numerical resolution, or that the coarsest numerical 
resolution we chose seems to be enough to properly describe the problem.

It is very likely that the collapse is a numerical artefact triggered by the 
boundary conditions at the outer radial boundary. Since we used ``outflow'' 
conditions, all gradients are set to zero at the boundary. This choice affects 
the radial component of the Lorentz force
\begin{equation}
F_{R} = -\frac{1}{2}\,\frac{\partial\,B_\phi^2}{\partial\,R} - \frac{B_\phi^2}{R}
-\frac{1}{2}\,\frac{\partial\,B_Z^2}{\partial\,R} +B_Z\,
\frac{\partial\,B_R}{\partial\,Z} \,,
\end{equation}
i.e. it cancels the radial gradient of the magnetic pressure, but it leaves 
the pinching force (second term on the right-hand side) unaffected. This may 
result in artificial collimation of the flow \citep[e.g.][]{UKR99,ZFR07}. Only 
in model ER2, where the toroidal magnetic field component $B_\phi$ (not only 
the density and the $Z$ velocity component) is damped to zero at the outer 
radial boundary, the collapse can be avoided. Although the simulation time 
is very short ($3.6\,t_0$), we can see that the model is stationary up to this 
point, after a short period of expansion due to higher thermal and magnetic 
pressure in the inner region which is also present in the science simulations 
(see below).  

Other effects caused by the very small values of $\rho$ and $v_Z$ in the 
external region as in model CD5 may be present. The fact, whether the outer 
radial boundary is causally connected with the jet from the very beginning or 
not, seems to be irrelevant for the presence or absence of the collapse.

In conclusion, it seems to be imperative that either the toroidal magnetic field
component $B_\phi$ should be very small at the outer radial boundary or that all
quantities should be modified {\em self-consistently} in order to maintain an 
equilibrium in the external region.

In all test models, a practical problem occurs: the time step decreases by 
three up to nine orders of magnitudes, making it impossible to further advance 
the simulations. This happens in the collapsing models as well as in model ER2.
In the latter, all quantities are exponentially damped to very small values 
close to the outer radial boundary, the large gradient in the last grid cell 
meets the zero gradient boundary condition in the ghost cells, which triggers 
leaking of matter into the domain and creating large velocities of all waves 
close to the boundary. This results in very small timesteps. In order to 
continue our studies and start a parameter study we are thus forced to 
``truncate the truncation'' as described in the next section.

\section{Parameter study of steady solutions} \label{sec_science_runs}

\subsection{Initial conditions}

In the science models SC1a -- SC5, we use the following approach: we modify 
again all quantities in one of the two components of the outflow, and second, 
we initialize the external region with another analytical solution with 
slightly changed parameters. One can show that if 
$\left[ \rho\,(R,Z) \,, p\,(R,Z) \,, \vec v\,(R,Z) \,, \vec B\,(R,Z) \right]$
is a solution of the MHD equations then  
$\left[ \rho\,'\,(R\,',Z\,') \,, p\,'\,(R\,',Z\,') \,, \vec v\,'\,(R\,',Z\,') 
\,, \vec B\,'\,(R\,',Z\,') \right]$ with
\begin{eqnarray} \label{scalings1}
R\,' = \lambda_1\,R \,, &\qquad& Z\,' = \lambda_1\,Z \,, \nonumber \\
\vec B\,' = \lambda_2\,\vec B \,, &\qquad&
\vec v\,' = \sqrt{\frac{\lambda_3}{\lambda_1}}\,\vec v \,, \nonumber \\
\rho\,' = \frac{\lambda_1\,\lambda_2^2}{\lambda_3}\,\rho \,, &\qquad&
p\,' = \lambda_2^2\,p \,, \nonumber \\
\mathcal{M}\,' &=& \lambda_3\,\mathcal{M}
\end{eqnarray}
is also a solution of the same set of equations. These transformations have 
also implications for the integrals of motion
\begin{eqnarray} \label{scalings2}
\Psi_A\,' (A) = \lambda_2\,\sqrt{\frac{\lambda_1}{\lambda_3}}\,\Psi_A\, (A)
\,, &\qquad& \Omega\,' (A) = \sqrt{\frac{\lambda_3}{\lambda_1^3}}\,\Omega (A) 
\,, \nonumber \\
L\,' (A) = \sqrt{\lambda_1\,\lambda_3}\,L (A) \,, &\qquad&
E\,' (A) = \frac{\lambda_3}{\lambda_1}\,E (A) \,, \nonumber \\
Q\,' (A) &=& \lambda_2^{2-2\,\Gamma}\,\left( \frac{\lambda_3}{\lambda_1} 
\right)^\Gamma\,Q (A) \,.
\end{eqnarray}
Since in our case, both solutions should have the same central object, we set
$\lambda_3 = 1$. We also ignore the scaling of $R$ and $Z$, i.e. we do not 
initialize e.g. $\rho\,'\,( R\,', Z\,' )$, but $\rho\,'\,( R, Z )$ in the outer 
region. Formally, this is not a solution of the MHD equations, since the gravity
term explicitly depends on the length scale. In our case, however, the 
gravitational force is small compared to the other forces, thus this slight 
inconsistency is unimportant. 

We match both solutions by using the function 
\begin{equation}
\mathcal{Q} = \mathcal{Q}_{\rm jet}\,\exp [- (\alpha/\alpha_{\rm trunc})^2 ] + 
\mathcal{Q}_{\rm ext}\, \left( 1 - \exp [- (\alpha/\alpha_{\rm trunc})^2 ] 
\right)
\end{equation}
for all quantities. 

Note that model ER2 can be also described by this ansatz, when we set 
$\mathcal{Q}_{\rm ext} = 0$ , i.e. simply damp all quantities with the 
exponential factor, or equivalently if we set $\lambda_1 \to \infty$ and 
$\lambda_2 \to 0$. However, this truncation led to problems with the 
time step (see Sec.~\ref{sec_testsims}). In model SC1a, we try to mimic this 
behavior less drastically by setting $\lambda_1 = 10^3$ and 
$\lambda_2 = 10^{-3}$ ($v\,' = 10^{-3/2}\,v$, $B\,' = 10^{-3}\,B$, 
$p\,' = 10^{-6}\,p$, $\rho\,' = 10^{-3}\,\rho$). 

In Eqs.~(\ref{scalings1})--(\ref{scalings2}), several special cases can be 
distinguished, coinciding with combinations of $\lambda_1$ and $\lambda_2$ where
quantities or integrals remain unchanged in both solutions: 
\begin{enumerate}
\item If $\lambda_1 = 1$, $R$, $Z$, $\vec v$, $\Omega (A)$, $L (A)$ and $E (A)$
remain the same.
\item If $\lambda_2 = 1$, $\vec B$ and $p$ are the same.
\item If $\sqrt{\lambda_1}\,\lambda_2 = 1$, $\rho$ and $\Psi_A\, (A)$ remain 
the same.
\end{enumerate}
The additional case, where $\lambda_1 = \lambda_2 = 1$, is of course 
identical with the unchanged analytical solution of V00 and 
therefore uninteresting in this study. For each choice of $\lambda_1$ and 
$\lambda_2$ we can create two different models, depending on whether the 
unchanged solution is inside or outside the solution with primed quantities.

In model SC2, we use the scaling parameters $\lambda_1 = 100$ and 
$\lambda_2 = 0.1$ (case 3.), i.e $v\,' = 0.1\,v$, $B\,' = 0.1\,B$, 
$p\,' = 0.01\,p$, $\rho\,' = \rho$, in the exterior region. In model SC3 we use 
the same scaling parameters, but apply the primed solution in the interior. In 
model SC4, our combination is $\lambda_1 = 1$ and $\lambda_2 = 0.1$ (case 1.) 
in the exterior and in model SC5 in the interior, leading to $v\,' = v$, 
$B\,' = 0.1\,B$, $p\,' = 0.01\,p$, $\rho\,' = 0.01\,\rho$.

\begin{table*}
\caption{List of numerical science models}
\label{tbl_models}
\centering
\begin{tabular}{l l l l l}
\hline\hline
Name & Grid size & Resolution & $t_{\rm end}$ & Description \\
\hline
SC1a & [0,50] $\times$ [6,100]  & 200 $\times$ 400 & 250.0 & 
$\alpha_{\rm trunc} = 0.4$, external analytical solution $\lambda_1 = 10^3$,
$\lambda_2 = 10^{-3}$ \\
SC1b & [0,50] $\times$ [6,100]  & 200 $\times$ 400 & 250.0 & 
$\alpha_{\rm trunc} = 0.2$, external analytical solution $\lambda_1 = 10^3$,
$\lambda_2 = 10^{-3}$ \\
SC1c & [0,50] $\times$ [6,100]  & 200 $\times$ 400 & 250.0 & 
$\alpha_{\rm trunc} = 0.1$, external analytical solution $\lambda_1 = 10^3$,
$\lambda_2 = 10^{-3}$ \\
SC1d & [0,50] $\times$ [6,100]  & 200 $\times$ 400 & 250.0 & 
$\alpha_{\rm trunc} = 0.01$, external analytical solution $\lambda_1 = 10^3$,
$\lambda_2 = 10^{-3}$ \\
SC1e & [0,50] $\times$ [6,100]  & 200 $\times$ 400 & 250.0 & 
$\alpha_{\rm trunc} = 0.001$, external analytical solution $\lambda_1 = 10^3$,
$\lambda_2 = 10^{-3}$ \\
SC2     & [0,50] $\times$ [6,100]  & 200 $\times$ 400 & 250.0 & 
$\alpha_{\rm trunc} = 0.4$, external analytical solution $\lambda_1 = 100$, 
$\lambda_2 = 0.1$ \\
SC3     & [0,50] $\times$ [6,100]  & 200 $\times$ 400 & 250.0 & 
same as model SC2, but solutions are swapped \\
SC4     & [0,50] $\times$ [6,100]  & 200 $\times$ 400 & 250.0 & 
$\alpha_{\rm trunc} = 0.4$, external analytical solution $\lambda_1 = 1$, 
$\lambda_2 = 0.1$ \\
SC5     & [0,50] $\times$ [6,100]  & 200 $\times$ 400 & 250.0 & 
same as model SC4, but solutions are swapped \\
\hline
\end{tabular}
\end{table*}

\subsection{Results of the science simulations}

\subsubsection{Structure of the flow}

As expected a priori, the flows behave qualitatively very differently, depending
on whether the scaled-down solution is inside or outside the original one. 

\begin{figure*}[ht!]
  \centering
  \includegraphics[width=0.8\textwidth]{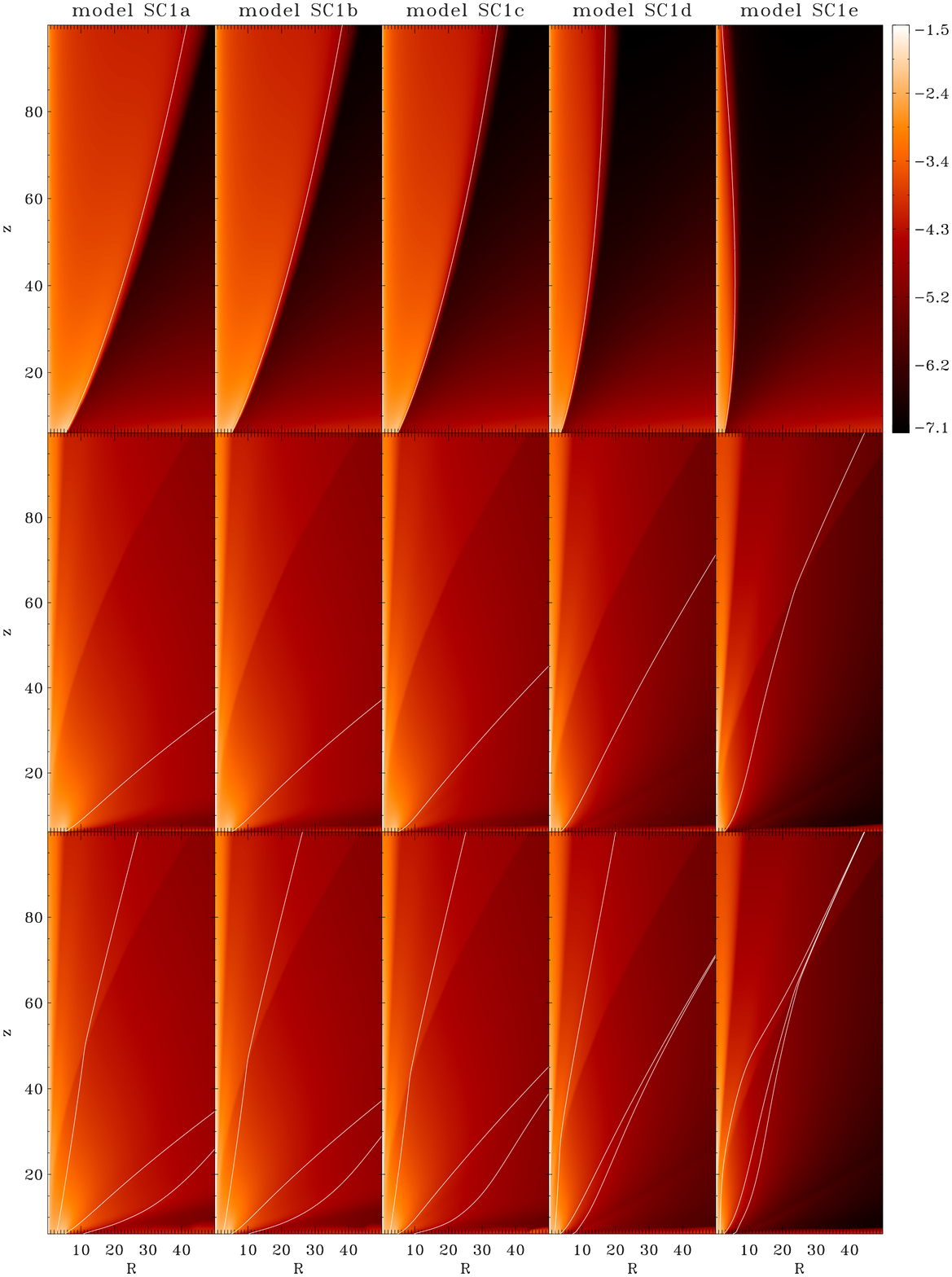}
  \caption{Structure of the flow (logarithmic density plots) for models 
    SC1a -- SC1e (from left to right) at timesteps $t = 0$ (top), $t = 25$ 
    (middle) and $t = 50\,t_0$ (bottom), respectively. Also plotted is the 
    magnetic field line anchored in the lower boundary where 
    $\alpha = \alpha_{\rm trunc}$ (white line)). At $t = 50\,t_0$, we also plot 
    the two field lines with half and twice the radius of that of the 
    truncation field line which are used in Sec.~\ref{sec_int}, Fig. 
    \ref{Fig_quantities_fl}}
  \label{Fig_struct_modelsSC1}
\end{figure*}

\begin{figure*}[ht!]
  \centering
  \includegraphics[width=0.8\textwidth]{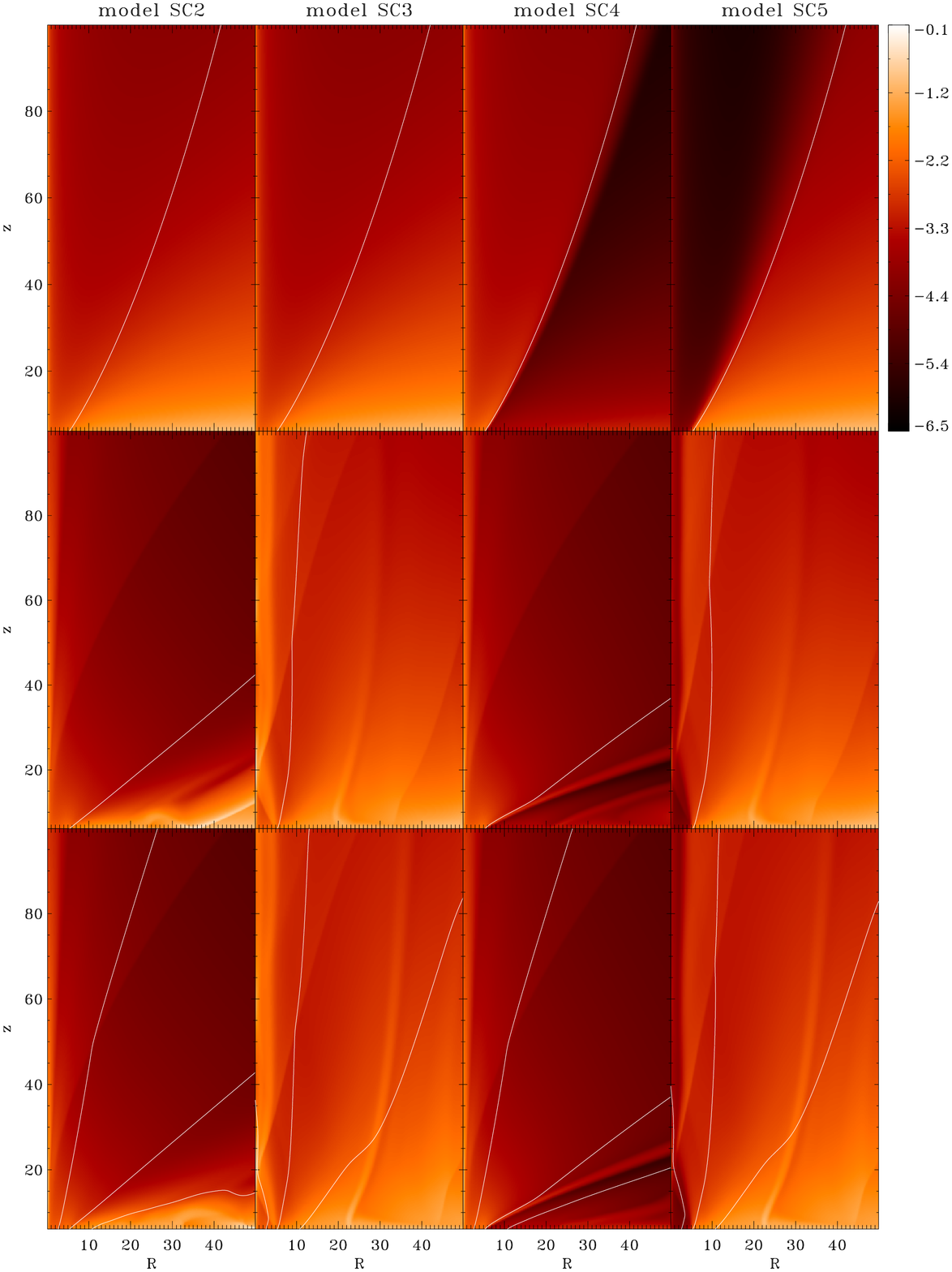}
  \caption{Structure of the flow (logarithmic density plots) for models 
    SC2 -- SC5 (from left to right) at timesteps $t = 0$ (top), $t = 25$ 
    (middle) and $t = 50\,t_0$ (bottom), respectively. Also plotted is the 
    magnetic field line anchored in the lower boundary where 
    $\alpha = \alpha_{\rm trunc}$ (white line). At $t = 50\,t_0$, we also plot 
    the two field lines with half and twice the radius of that of the 
    truncation field line which are used in Sec.~\ref{sec_int}, Fig. 
    \ref{Fig_quantities_fl}}
  \label{Fig_struct_modelsSC2-5}
\end{figure*}

In the cases where the quantities are scaled down in the exterior region, the 
opening angle of the flow increases, see Figs.~\ref{Fig_struct_modelsSC1} 
and~\ref{Fig_struct_modelsSC2-5}). The jet surface is 
pushed outwards by the higher thermal and magnetic pressure in the inner 
region, which also dilutes while expanding. A new equilibrium is 
established within several $t_0$ and is stable for at least 250 $t_0$. The 
final opening angles seen in the density plots seem to be too 
large for calling the flow a collimated jet. Using the truncation field line 
plotted in the figures, we find angles of about 40$^\circ$--50$^\circ$ in models 
SC1a -- SC2 and model SC4. In paper II of this series, we assess the 
``apparent'' opening angle by calculating emission maps and find that the 
emitting region is very collimated in all models (the opening angles are of the 
order of 10$^\circ$--20$^\circ$).

If all quantities are reduced in the internal region (models SC3 and SC5), then 
the opening angle decreases. Again a new equilibrium is established which is 
stable for at least 250 $t_0$, see Fig.~\ref{Fig_struct_modelsSC2-5}. The final 
opening angles in these runs are very small, around 5$^\circ$. New features are 
several shocks, the innermost of which is also collimated following magnetic 
field lines.

The expansion of the flow in models SC1a--SC1e, SC2 and SC4 
(Fig.~\ref{Fig_struct_modelsSC1}), and the collimation in models SC3 and SC5 
(Fig.~\ref{Fig_struct_modelsSC2-5}) is clearly visible in the evolution of the 
jet radius. While the jet radius is very similar in both models showing the 
collimation, a clear trend can be identified in the first case: the largest 
expansion occurs in model SC4, followed by the models SC1a, SC1b, SC2, SC1c, 
SC1d and SC1e. Surprisingly model SC1a, which is the model with the smallest 
values of thermal pressure and magnetic field in the outer region, is not the 
one with the highest expansion. Although models SC2 and SC4 have the same 
scalings for thermal pressure and magnetic field, they show different degrees of
expansion.

In models SC3 and SC5, again two shocks are present which move outwards. When 
the inner region shrinks due to its lower thermal and magnetic pressure compared
to the exterior, two waves propagate towards larger radii emitted from the 
truncation field line. One wave travels faster than the other, and since both 
waves compress the medium, they seem to be again a slow- and a 
fast-magnetosonic wave. The fast wave has reached the outer radial boundary at 
50 $t_0$, only its part at lower latitudes ($z<20$) is still visible inside the 
domain. The slow wave stops when its lower part (at $z = 6$) reaches $R = 22$, 
which is exactly the location where the slow-magnetosonic critical surface 
crosses the lower boundary. Since the slow-magnetosonic wave cannot cross the 
corresponding critical surface, it is then attached to this point and develops a
standing and steepening shock inside the domain. A third shock, which is also 
present in the model ADO, develops at the  fast magnetosonic separatrix surface 
(FMSS). This weak shock acts as a "wall"  protecting the sub-fast flow from the 
imposed modifications close to the axis (GVT06, M08). Note that the poloidal 
field lines develop a bend at the position of all shocks.   

In models SC1a and SC1b, a small dip is present after about 70 $t_0$ and 150 
$t_0$, respectively. Before this dip, the radii are almost constant, as well as 
in the other models (Fig.~\ref{BCeffects_jetradii}, top, dashed line). 

\subsubsection{Are there boundary effects in the science runs?}

We have also run models SC1a', SC3' and SC4' with a larger 
domain ([0,100]$\times$[6,200]) to check whether the science runs presented 
above are affected by the boundaries. 

\begin{figure}[ht!]
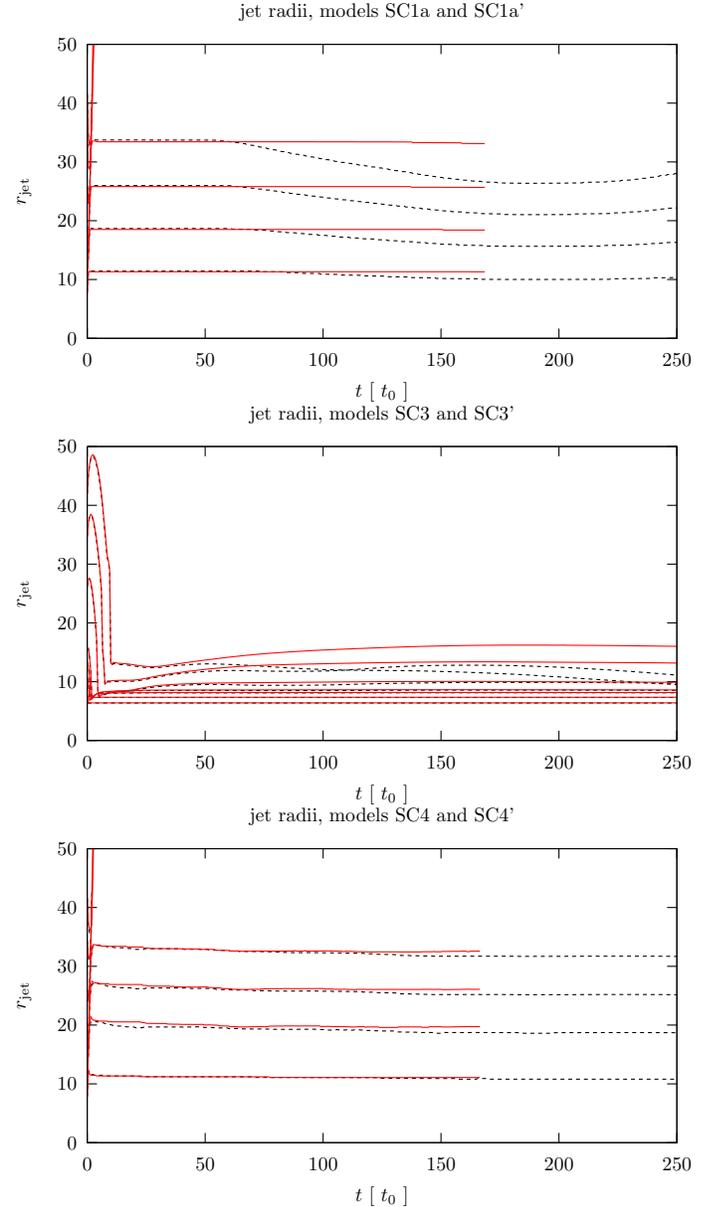

  \centering
  \includegraphics[width=\columnwidth]{fig6a.eps}
  \includegraphics[width=\columnwidth]{fig6b.eps}
  \includegraphics[width=\columnwidth]{fig6c.eps}
  \caption{Comparison of the jet radius evolution in models SC1a and 
    SC1a', models SC3 and SC3' and models SC4 and SC4' 
    (black dashed: unprimed models, red solid: primed models)}
  \label{BCeffects_jetradii}
\end{figure}

The results concerning the jet radii measured with the truncation field line are
only changed by a few percent in models SC1a' (with respect to the 
constant part in model SC1a) and SC4', with less effects in model 
SC4' than in model SC1a', see Fig.~\ref{BCeffects_jetradii}. The 
small dip seen in model SC1a is smoothed out, thus it was indeed created by 
boundary effects as suspected. In model SC3', larger changes exist at 
higher $z$ values ($z > 75$) after about 50 $t_0$.  While in model SC3, the 
radii decrease, they stay constant in model SC3'. 

Therefore, throughout the remainder of the paper, the results of all runs at 50 
$t_0$ will be considered as ``real'' final stationary states of the solutions.

\subsubsection{Force balance along the jet boundary}

\begin{figure}[ht!]
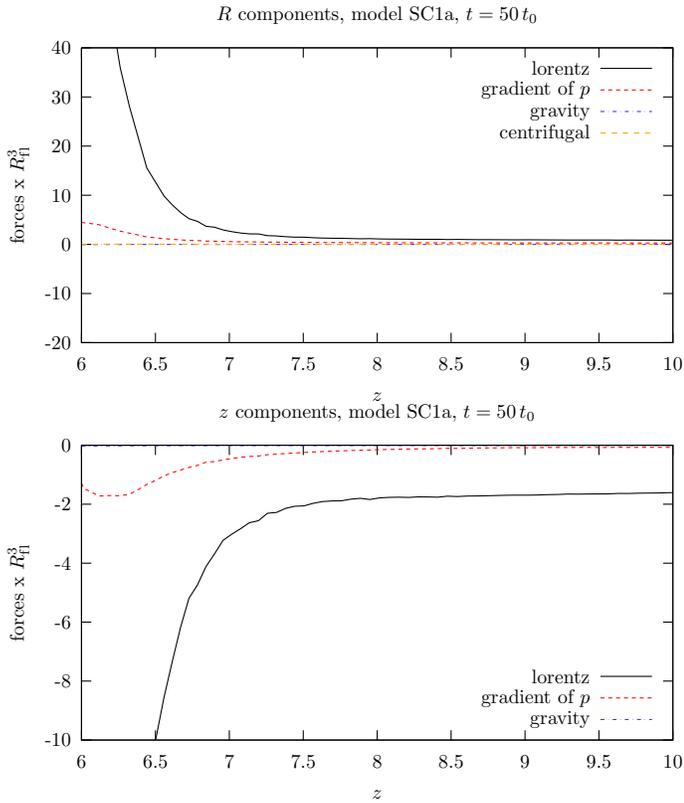

  \includegraphics[width=\columnwidth]{fig7a.eps}
  \includegraphics[width=\columnwidth]{fig7b.eps}
  \caption{$R$ (top) and $z$ (bottom) components of forces along the jet 
    boundary at the final simulated timestep (at 50 $t_0$) in models SC1a.
    The components are measured at the truncation field line plotted in 
    previous figures and multiplied by $R_{\rm fl}^3$ with $R_{\rm fl}$ the local
    cylindrical radius of the field line}
  \label{Fig_Forces_modelSC1a}
\end{figure}

The interplay between the $R$ components of the pressure gradient
and the Lorentz force (which, as expected, always dominates in those radially 
self-similar disk-wind models) is responsible for collimating the flow or 
triggering its expansion in all models (see Figs.~\ref{Fig_Forces_modelSC1a} -- 
\ref{Fig_Forces_modelSC3}). In the expanding models SC1a--SC1e and SC2, the 
Lorentz force and the pressure gradient are directed outwards, while they are 
directed inwards in the models SC3 and SC5. The first is also true for model 
SC4, the pressure gradient turns inwards between $z = 7$ and $z = 14$. 
Furthermore, in all models except for SC4, the pressure gradient is only 
important at lower $z$ values between 6 -- 8.

\begin{figure}[ht!]
  \includegraphics[width=\columnwidth]{fig8a.eps}
  \includegraphics[width=\columnwidth]{fig8b.eps}
  \caption{Same as in Fig.~\ref{Fig_Forces_modelSC1a}, but for model SC3.}
  \label{Fig_Forces_modelSC3}
\end{figure}

\subsubsection{Quantities at the outer axial boundary and the properties of 
the jet}

In Figs. \ref{Fig_quantities_modelSC1a}--\ref{Fig_quantities_modelSC5}, all 
eight magnetohydrodynamic quantities are plotted along the outer axial boundary 
at $z = 100$ for our models SC1a -- SC5. In each panel, we show the 
initial profile of our model (solid line), the profile of the unchanged 
analytical solution (dash-dotted line) and the final profile at $t = 50\,t_0$ 
(dashed line). One can clearly see how the initial profile follows the 
analytical solution in the interior region and is damped in the exterior in 
models SC1a--SC2 and SC4 and vice versa in models SC3 and SC5.

In models SC1a--SC1e, SC2 and SC4, at large radii, most profiles have a 
similar shape as the unchanged analytical solution, we started with. The 
original break is mostly leveled out in all of the profiles. However, at small 
radii ($R < 7$) the final solution deviates drastically. This is expected, 
since the analytical solution is modified close to the axis to avoid the 
intrinsic singularities. In density, pressure, axial ($z$) and toroidal 
($\phi$) velocity components, as well as in axial magnetic field, a peak
is present where all quantities are higher than at larger radii ($R > 7$). The 
radial ($R$) velocity and magnetic field components and the toroidal magnetic 
field on the other hand show strong deviations from the initial profiles with 
strong declines close to the axis in the former two quantities and a steep rise
and a minimum at about $R = 4$ in the latter. The relative height of the 
central peak, the height of the remaining break and the depth of the minimum 
in $B_{\phi}$ are different across the five models. Although the models SC2 
and SC4 are special cases, where the density and the velocity, respectively, 
is unchanged in the inner and outer region, the final profiles in all models 
including these two are similar.

In models SC3 and SC5, the initial profiles have to be changed dramatically 
to stop the decrease of the opening angle and the collapse of the jet. The 
profiles are much more complicated and less smooth than in the other five cases
described above. Again some final profiles show a similar shape as the 
unchanged analytical solution at large radii. However, superimposed on all of 
them is a shock structure with jumps in density, pressure, radial  magnetic 
field and toroidal velocity (at about $R = 4$, $R = 13$, $R = 25$, and $R = 37$
in model SC3). The radial magnetic field changes sign at the innermost shock 
where the density and pressure rise steeply and a minimum in axial velocity is 
present in model SC3. In model SC5, the radial magnetic field also changes sign 
at the innermost shock which is, however, now characterized by a steep decline 
in density and a steep rise in axial velocity.

 \begin{figure}[ht!]
  \includegraphics[width=0.45\columnwidth]{fig10a.eps}
  \includegraphics[width=0.45\columnwidth]{fig10b.eps}

  \includegraphics[width=0.45\columnwidth]{fig10c.eps}
  \includegraphics[width=0.45\columnwidth]{fig10d.eps}

  \includegraphics[width=0.45\columnwidth]{fig10e.eps}
  \includegraphics[width=0.45\columnwidth]{fig10f.eps}

  \includegraphics[width=0.45\columnwidth]{fig10g.eps}
  \includegraphics[width=0.45\columnwidth]{fig10h.eps}
  \caption{Quantities at the outer axial boundary in model SC1a: density (top 
  left), pressure (top right), velocity components in $R$ (second row, left), 
  $z$ (second row, right) and $\phi$ (third row, left), magnetic field 
  components in $R$ (third row, right), $z$ (bottom left) and $\phi$ (bottom 
  right); given are the initial profiles (solid), the final profiles at 
  $50\,t_0$ (dashed) and the profiles in the unchanged analytical solution of 
  V00.}
  \label{Fig_quantities_modelSC1a}
\end{figure} 

\begin{figure}[ht!]
  \includegraphics[width=0.45\columnwidth]{fig15a.eps}
  \includegraphics[width=0.45\columnwidth]{fig15b.eps}

  \includegraphics[width=0.45\columnwidth]{fig15c.eps}
  \includegraphics[width=0.45\columnwidth]{fig15d.eps}

  \includegraphics[width=0.45\columnwidth]{fig15e.eps}
  \includegraphics[width=0.45\columnwidth]{fig15f.eps}

  \includegraphics[width=0.45\columnwidth]{fig15g.eps}
  \includegraphics[width=0.45\columnwidth]{fig15h.eps}
  \caption{Same as in Fig.~\ref{Fig_quantities_modelSC1a}, but for the model 
    SC2.}
  \label{Fig_quantities_modelSC2}
\end{figure} 

\begin{figure}[ht!]
  \includegraphics[width=0.45\columnwidth]{fig16a.eps}
  \includegraphics[width=0.45\columnwidth]{fig16b.eps}

  \includegraphics[width=0.45\columnwidth]{fig16c.eps}
  \includegraphics[width=0.45\columnwidth]{fig16d.eps}

  \includegraphics[width=0.45\columnwidth]{fig16e.eps}
  \includegraphics[width=0.45\columnwidth]{fig16f.eps}

  \includegraphics[width=0.45\columnwidth]{fig16g.eps}
  \includegraphics[width=0.45\columnwidth]{fig16h.eps}
  \caption{Same as in Fig.~\ref{Fig_quantities_modelSC1a}, but for the model 
    SC3.}
  \label{Fig_quantities_modelSC3}
\end{figure} 

\begin{figure}[ht!]
  \includegraphics[width=0.45\columnwidth]{fig17a.eps}
  \includegraphics[width=0.45\columnwidth]{fig17b.eps}

  \includegraphics[width=0.45\columnwidth]{fig17c.eps}
  \includegraphics[width=0.45\columnwidth]{fig17d.eps}

  \includegraphics[width=0.45\columnwidth]{fig17e.eps}
  \includegraphics[width=0.45\columnwidth]{fig17f.eps}

  \includegraphics[width=0.45\columnwidth]{fig17g.eps}
  \includegraphics[width=0.45\columnwidth]{fig17h.eps}
  \caption{Same as in Fig.~\ref{Fig_quantities_modelSC1a}, but for the model 
    SC4.}
  \label{Fig_quantities_modelSC4}
\end{figure} 

\begin{figure}[ht!]
  \includegraphics[width=0.45\columnwidth]{fig18a.eps}
  \includegraphics[width=0.45\columnwidth]{fig18b.eps}

  \includegraphics[width=0.45\columnwidth]{fig18c.eps}
  \includegraphics[width=0.45\columnwidth]{fig18d.eps}

  \includegraphics[width=0.45\columnwidth]{fig18e.eps}
  \includegraphics[width=0.45\columnwidth]{fig18f.eps}

  \includegraphics[width=0.45\columnwidth]{fig18g.eps}
  \includegraphics[width=0.45\columnwidth]{fig18h.eps}
  \caption{Same as in Fig.~\ref{Fig_quantities_modelSC1a}, but for the model 
    SC5.}
  \label{Fig_quantities_modelSC5}
\end{figure} 

\subsubsection{The integrals of motion} \label{sec_int}
 
\begin{figure*}[ht!]
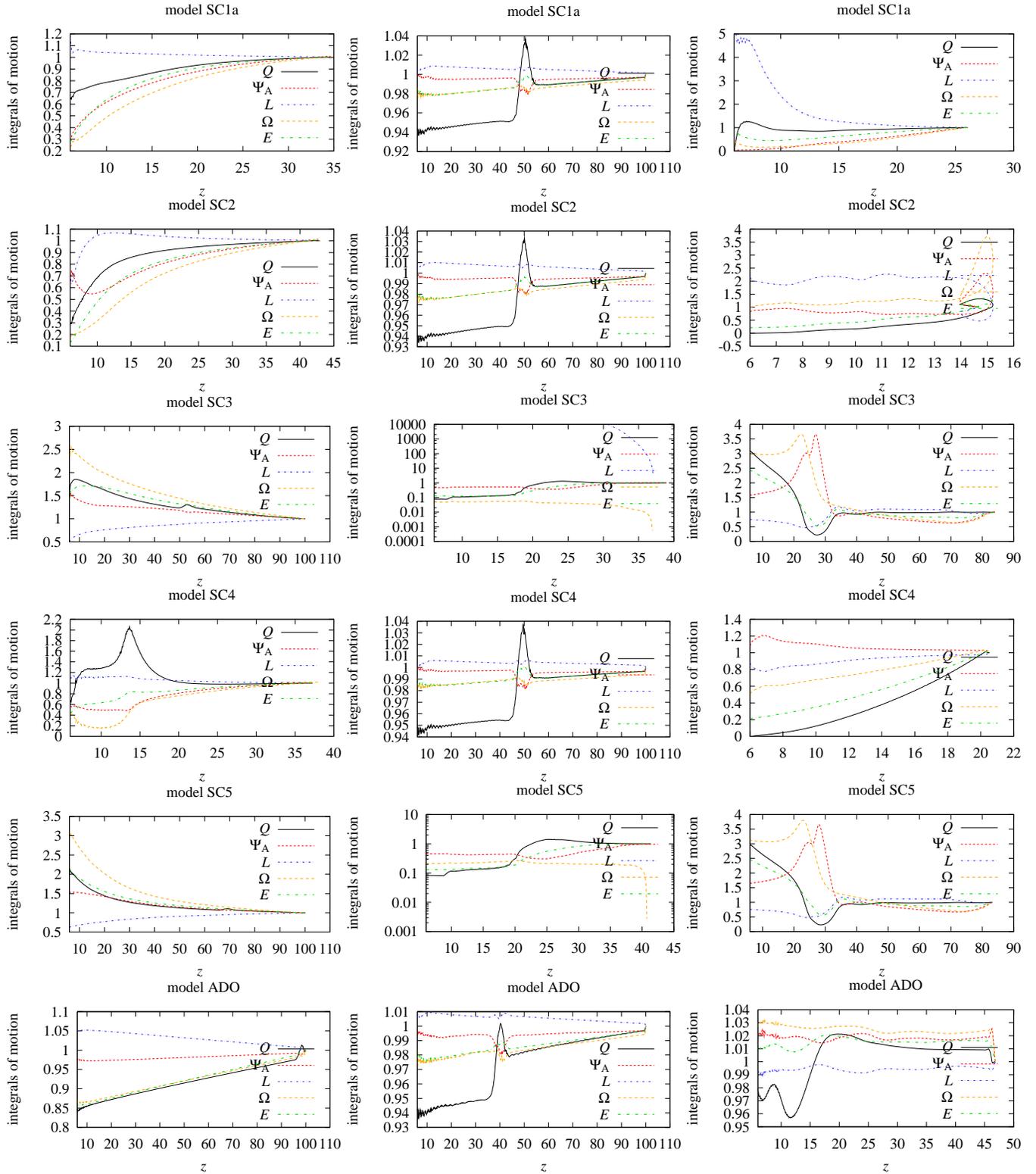

  \includegraphics[width=0.32\textwidth]{fig19a.eps}
  \includegraphics[width=0.32\textwidth]{fig19b.eps}
  \includegraphics[width=0.32\textwidth]{fig19c.eps}

  \includegraphics[width=0.32\textwidth]{fig20a.eps}
  \includegraphics[width=0.32\textwidth]{fig20b.eps}
  \includegraphics[width=0.32\textwidth]{fig20c.eps}

  \includegraphics[width=0.32\textwidth]{fig20d.eps}
  \includegraphics[width=0.32\textwidth]{fig20e.eps}
  \includegraphics[width=0.32\textwidth]{fig20f.eps}

  \includegraphics[width=0.32\textwidth]{fig20g.eps}
  \includegraphics[width=0.32\textwidth]{fig20h.eps}
  \includegraphics[width=0.32\textwidth]{fig20i.eps}

  \includegraphics[width=0.32\textwidth]{fig20j.eps}
  \includegraphics[width=0.32\textwidth]{fig20k.eps}
  \includegraphics[width=0.32\textwidth]{fig20l.eps}

  \includegraphics[width=0.32\textwidth]{fig20m.eps}
  \includegraphics[width=0.32\textwidth]{fig20n.eps}
  \includegraphics[width=0.32\textwidth]{fig20o.eps}
  \caption{Integrals of motion $\Psi_A\, (A)$, $\Omega (A)$, $L (A)$, $E (A)$
  and $Q (A)$, normalized to their values at their end point, where they leave 
  the domain, along the truncation field line (left), along a second field line 
  anchored at half of the radius (middle) and along a third field line anchored
  at twice of the radius (right) for models SC1a, SC2 -- SC5 and ADO (from top 
  to bottom).}
  \label{Fig_quantities_fl}
\end{figure*}

In Fig. \ref{Fig_quantities_fl}, we plot the
integrals of motion given by Eqs.~(\ref{integrals1}) -- (\ref{integrals2}), 
along the truncation field line at the interface of both regions, along an inner
field line which is anchored at half of the radius of the truncation field line,
and along an outer field line anchored at twice the radius. 

When we compare the integrals along the inner field line in the models 
SC1a -- SC1e, SC2 and SC4 with those in the analytical model ADO, we can see 
that the flow in the inner region is very similar to the analytical solution we 
started with. An exception is model SC1e, where the inner field line is already 
very close to the symmetry axis, i.e. our modifications there affect the 
integrals of motion. All integrals of motion converge smoothly to an asymptotic 
value. The deviation from this value are within 6 \% for $z$ values above 20. 

In the remaining models SC3 and SC5, the behavior of the integrals of motion 
along the inner field line is very different to the other models, but similar 
to each other. At about $z = 50$, peaks and dips are visible in all integrals,
however,  with the most pronounced in $Q$. 

In the outer region, all integrals seem to converge in models SC1a--SC1e, 
SC2, SC3 and SC5. The turnovers in model SC2 are a result of a turnover in the 
field line itself. In model SC4, $L$ and $\Psi_A$ have already converged, the 
other integrals still vary. In conclusion, in most of our models also the 
external region reaches a steady state.

\subsubsection{Topology of the current lines}

\begin{figure}[htb!]
  \centering
  \includegraphics[width=\columnwidth]{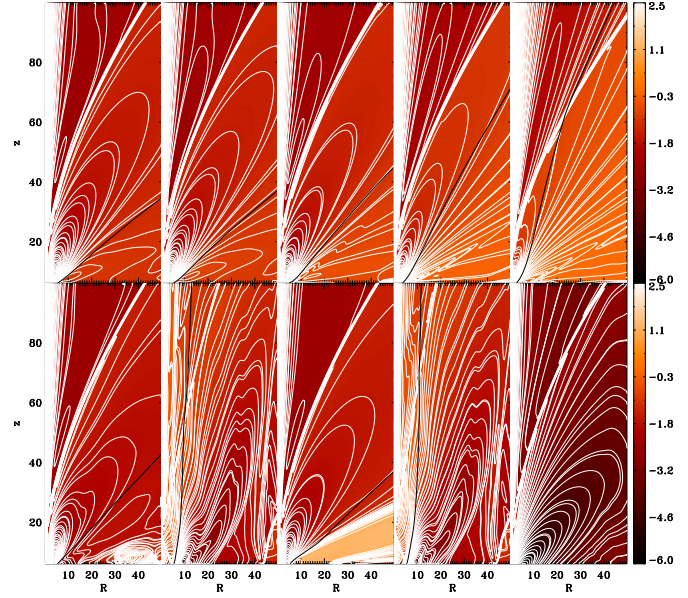}
  \caption{Poloidal currents $R\,B_\phi = const$ of models 
  SC1a--SC1e (top) and models SC2--SC5 and ADO (bottom) at the final timestep 
  (at 50 $t_0$) are plotted.}
  \label{Fig_current}
\end{figure}

In Fig. \ref{Fig_current}, we plot the poloidal currents for our nine 
science models SC1a -- SC5. In model ADO, as well as in the inner region of 
models SC1a--SC2 and SC4, which remains unchanged with respect to the 
analytical solution, a re-adjusted FMSS
is visible as shock in the density plots (Figs. \ref{Fig_struct_modelsSC1} --
\ref{Fig_struct_modelsSC2-5}, cf. GVT06 and M08). The currents in these 
models are counterclockwise upstream and downstream of the FMSS. Only along
the FMSS the current close, i.e. the lines are clockwise. This distribution of 
currents, and in particular the direction of the resulting 
$\vec J_{\rm p} \times \vec B_{\phi}$ force, is consistent with the 
decollimation and deceleration that the flow experiences as it passes through 
the shock. Thus, one of the effects of the new FMSS is to bend the streamlines 
away from the z-axis avoiding the overcollimation property of the original 
analytical solution. The collimation and decollimation processes that can be 
derived from such a plot are also discussed in GVT06. Towards the 
outer region, the morphology of the current lines is distorted and of 
completely different topology. In models SC4 and SC1a--SC1e, the outer region 
is separated by a second shock with a current sheet from the inner region.

In models SC3 and SC5, the FMSS is also present. Again at the shock, a current 
sheet separates two region with counterclockwise currents. Furthermore, closer 
to the rotation axis, another current sheet forms an X shape with the FMSS at 
$(R, z) = (5, 30)$ and a third current sheet in the lower right corner of the 
domain develops. 

\section{Summary}

In this paper, we have studied the effects of imposing an outer radius of the 
underlying accreting disk, and thus also of the outflow, on the topology, 
structure and time-dependence of a well studied radially self-similar 
analytical solution (V00). We matched an unchanged and a scaled-down analytical 
solution of V00 and found in all these cases steady two-component solutions. 

We showed that the boundary between both solutions is always shifted towards 
the solution with reduced quantities. Especially, the reduced thermal and 
magnetic pressure change the perpendicular force balance at the ``surface''
of the flow. In the models where the scaled-down analytical solution is
{\em outside} the unchanged one, the inside solution converges to another
analytical solution with different parameters. In the models where the 
scaled-down analytical solution is {\em inside} the unchanged one, the whole 
two-component solution changes dramatically to support the flow from 
collapsing totally to the symmetry axis.

A result of the modification of the analytical solution at the symmetry axis 
is the formation of a weak shock at the fast magnetosonic separatrix surface 
(FMSS) which shields the flow upstream of the FMSS from the imposed 
modifications close to the axis. A result of the truncation of the analytical 
solution along some poloidal field line is the formation of two compressive MHD 
shocks: a fast shock that travels quickly downstream outside of the 
computational domain and a slow shock which as it travels downstream with lower 
speed, it is locked precisely at the position of the lower computational 
z-boundary where the analytical slow surface meets this boundary.  
 
Our truncated disk-wind solutions are stable for more than 50 $t_0$, i.e. 
several orbital periods at the truncation radius. These solutions may be 
relevant to describe observed jets, since the jet radii are too large in the 
untruncated analytic disk outflow solution, with respect to observed jet widths.
We also provide all quantities at the outer $z$ boundary which can now be used 
as boundary conditions for jet propagation studies.

In the following paper II, we calculate emission maps corresponding to such 
truncated disk-models and compare them with observations.

\begin{acknowledgements}
We acknowledge the improving comments and suggestions by the referee.
The present work was supported by the European Community's Marie 
Curie Actions - Human Resource and Mobility within the JETSET (Jet 
Simulations, Experiments and Theory) network under contract 
MRTN-CT-2004 005592.
\end{acknowledgements}

\end{document}